\documentclass[aps,prd,showpacs,nofootinbib]{revtex4}
\usepackage{graphicx}
\usepackage{amsfonts}
\usepackage{color}
\usepackage{amsmath}

\usepackage[final]{showkeys}

\newcommand{\cA}{\ensuremath{\mathcal{A}}}
\newcommand{\cB}{\ensuremath{\mathcal{B}}}

\newcommand{\cL}{\ensuremath{\mathcal{L}}}
\newcommand{\cM}{\ensuremath{\mathcal{M}}}

\newcommand{\cO}{\ensuremath{\mathcal{O}}}

\newcommand{\cU}{\ensuremath{\mathcal{U}}}
\newcommand{\cV}{\ensuremath{\mathcal{V}}}
\newcommand{\cW}{\ensuremath{\mathcal{W}}}

\newcommand*{\chpt}{\raise0.4ex\hbox{$\chi$}PT}
\newcommand*{\schpt}{S\raise0.4ex\hbox{$\chi$}PT}

\newcommand*{\etc}{\textit{etc.}}

\newcommand*{\Tr}{\operatorname{Tr}}

\newcommand{\ltwid}{\raise.3ex\hbox{$<$\kern-.75em\lower1ex\hbox{$\sim$}}}
\newcommand{\eq}[1]{Eq.~(\ref{eq:#1})}
\newcommand{\eqs}[2]{Eqs.~(\ref{eq:#1}) and (\ref{eq:#2})}
\newcommand{\eqsthru}[2] {Eqs.~(\ref{eq:#1}) through (\ref{eq:#2})}

\newcommand{\be}{\begin{equation}}
\newcommand{\ee}{\end{equation}}
\newcommand{\bea}{\begin{eqnarray}}
\newcommand{\eea}{\end{eqnarray}}
\newcommand{\beas}{\begin{eqnarray*}}
\newcommand{\eeas}{\end{eqnarray*}}

\begin{document}

\title{Unphysical phases in staggered chiral perturbation theory}
\author{Christopher~Aubin}\email{caubin@fordham.edu}
\affiliation{Department of Physics \& Engineering Physics,  Fordham University, Bronx, New York, NY 10458, USA}
\author{Katrina~Colletti}\email{kcolletti1@physics.tamu.edu}
\affiliation{Department of Physics \& Engineering Physics,  Fordham University, Bronx, New York, NY 10458, USA}
\author{George~Davila}\email{gdavila@fordham.edu}
\affiliation{Department of Physics \& Engineering Physics,  Fordham University, Bronx,
New York, NY 10458, USA}
\begin{abstract}

We study the phase diagram for staggered quarks using chiral perturbation theory. 
In 
beyond-the-standard-model simulations using a large number ($>8$) of staggered 
fermions, unphysical phases appear for coarse enough lattice spacing.  
We argue chiral perturbation theory can be used to interpret one of these phases.
In addition, we show only three 
broken phases for staggered quarks exist, at least 
for lattice spacings in the regime $a^2\ll \Lambda^2_{\rm QCD}$.

\end{abstract}
\pacs{11.15.Ha,11.30.Qc,12.39.Fe}

\maketitle

\section{Introduction}\label{sec:intro}

The fact that unphysical phases may arise in lattice simulations for coarse lattice spacings has been known for some time \cite{Aoki:1983qi,Sharpe:1998xm,Lee:1999zx,Aubin:2003mg,Aubin:2004dm}. Such phases arise when the squared mass of a meson becomes negative in a region of the relevant parameter space. When this occurs we must find the true minimum of the potential so that we can expand about the true ground state of the theory. Doing so for lattice simulations is important as the continuum limit cannot be properly taken unless they are performed in the unbroken, physical, phase, where the vacuum state has the symmetries of the action.

For staggered quarks, the case of interest here, unphysical phases appear when $ca^2  < - m$, where $m$ is the light quark mass, for some parameter $c$ (the specific form will be discussed in Sec.~\ref{sec:mass_deg}) arising from the $\cO(a^2)$ taste-symmetry breaking potential. This implies that these unphysical phases can be studied using rooted staggered chiral perturbation theory (r\schpt) \cite{Lee:1999zx,Aubin:2003mg}, which requires $a^2$ to be fine enough such that the low-energy effective theory is valid. Thus, we are interested in a region such that
\be\label{eq:condition}
	 \frac{m}{\Lambda_{\rm QCD}} \, \ltwid\,  a^2 \Lambda^2_{\rm QCD} \ll 1\ .
\ee
The first condition assumes the parameter $c<0$ and that $ca^2$ is large enough that one of the squared meson masses have become negative, 
while the second condition is necessary for our low-energy effective theory to be valid. 

If simulations are performed in the broken phase, one cannot use the numerical results to describe physical systems. As such, understanding where these unphysical phases occur and how to detect them is essential in understanding the system being simulated. In Ref.~\cite{Aubin:2004dm}, one unphysical phase for staggered quarks was studied and an analysis of the mass spectrum was performed, noting the possibility of additional broken phases in the system. However, it is clear that the phase in Ref.~\cite{Aubin:2004dm} is not seen in 2+1-flavor simulations (see Refs.~\cite{Aubin:2004fs,Bazavov:2009bb} for example).

In recent work looking into beyond-the-Standard-Model (BSM)
theories by Ref.~\cite{Cheng:2011ic} using 8 or 12 flavors of degenerate staggered quarks,\footnote{These would then be 4+4-flavor or 4+4+4-flavor simulations.} 
two broken phases were seen in additional to the standard physical phase.
One of the phases examined in Ref.~\cite{Cheng:2011ic} shares several features as the phase studied in Ref.~\cite{Aubin:2004dm}, as we discuss in this work.

In Ref.~\cite{Cheng:2011ic}, the authors found three distinct phases appearing in the staggered theory for 12 flavors of staggered quarks. The first phase, seen at weaker coupling, is the unbroken phase, as it retains the discrete shift symmetry of staggered fermiosn, and has the expected mass spectrum, at least approximately. The intermediate phase, at slightly stronger coupling, we argue falls within the window in \eq{condition} so that r\schpt\ is applicable, and is the broken phase seen in Ref.~\cite{Aubin:2004dm}. Finally, the phase that arises at the strongest coupling in Ref.~\cite{Cheng:2011ic} is outside the chiral regime, and thus cannot be studied using the methods of this paper.

One can use the replica method for r\schpt\ \cite{Bernard:2006zw} to generalize the results of Ref.~\cite{Aubin:2004dm} for $n_f$ degenerate flavors and $n_t$ tastes-per-flavor. We define $n_q \equiv n_f n_t$ as the number of quarks in our resulting theory. The phase studied in Ref.~\cite{Aubin:2004dm}, which we will refer to as the ``$A$-phase,'' appears when
\be\label{eq:Aphasereq}
	  a^2 \delta_{A-}' 
	  < - \frac{4}{3}(2\mu m + a^2 \Delta_A)\ ,
\ee
where $m$ is the light quark mass, and we are denoting $\delta_{A}'$, the hairpin term of Refs.~\cite{Aubin:2003mg,Aubin:2004dm}, as $\delta_{A-}'$. This is assuming three flavors of degenerate rooted-staggered quarks, but if we generalize this using the replica method to $n_f$ flavors and $n_t$ tastes, we can rewrite the condition for broken phase as
\begin{eqnarray}\label{eq:Aphasereq2}
	n_f n_t = n_q > 4 \left(\frac{2 \mu m + a^2\Delta_A}{-a^2\delta_{A-}'}\right)\ .
\end{eqnarray}

\begin{figure}[tb]
\begin{center}
\includegraphics[width=3in]{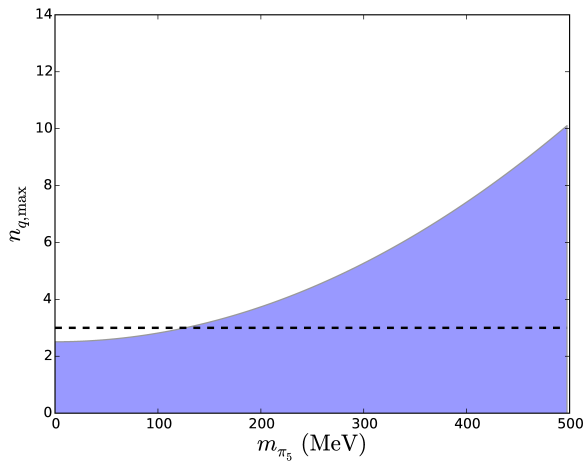}\\

\caption{The maximum allowed number of quarks as a function of pion mass to ensure a simulation is in the unbroken phase for the coarse asqtad MILC ensembles ($a\approx0.125$ fm). The dashed line is $n_q = 3$ and the shaded region shows allowed values of $n_q$ as a function of the pion mass. }
\label{fig:nqvsmpia}
\end{center}
\end{figure}

\begin{figure}[tb]
\begin{center}
\includegraphics[width=3in]{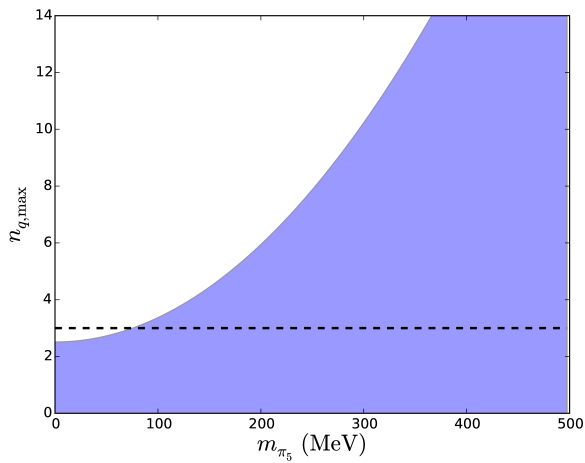}
\caption{The maximum allowed number of quarks as a function of pion mass to ensure a simulation is in the unbroken phase for the fine asqtad MILC ensembles ($a\approx0.09$ fm). The dashed line is $n_q = 3$ and the shaded region shows allowed values of $n_q$ as a function of the pion mass. }
\label{fig:nqvsmpib}
\end{center}
\end{figure}

Given a sample set of parameters in MILC simulations for the $a=0.125$ fm and $a=0.09$ fm asqtad ensembles for these values on the right-hand side \cite{MILCparams}, we show the maximum number of allowed quarks, $n_{q,\rm max}$, for the simulation to remain in the unbroken phase as a function of $m_{\pi_5}$ (the Goldstone pion mass) in Figs.~\ref{fig:nqvsmpia} and \ref{fig:nqvsmpib}. The shaded region shows allowed values of $n_q$ as a function of the pion mass.  The dashed lines in these figures indicate $n_q=3$, which is well below the limit for being in the unbroken phase (except for $m_\pi\,\ltwid\, 128$ MeV on the coarse ensemble and $m_\pi\,\ltwid\, 90$ MeV on the fine ensemble).

A simulation is more likely to be in the $A$-phase when we simulate 8 or 12 quarks than when we simulate fewer quarks. More specifically, if the (Goldstone) pion mass is 
around 500 MeV, the simulation would most likely be in the unbroken phase for 8 quarks (2 flavors, 4 tastes per flavor), while in the $A$-phase for 12 quarks.  Figures~\ref{fig:nqvsmpia} and \ref{fig:nqvsmpib} were generated using parameters from asqtad MILC ensembles for various lattice spacings. The specific picture will change with different staggered quarks such as nHYP staggered quarks \cite{Hasenfratz:2001hp,Hasenfratz:2007rf}, as are used in Ref.~\cite{Cheng:2011ic}, but qualitatively we would expect similar results.

In this paper we study the staggered phase diagram for all values of the r\schpt\ parameters that may arise during a simulation. 
In Ref.~\cite{Aubin:2004dm}, a third possible phase was discussed (which we will refer to as the $A'$-phase) and we show that it cannot occur. Instead, in addition 
to the $A$-phase, there are two other broken phases that we label the 
$V$-phase and the $T$-phase. 
We also show that one of the two broken phases seen in Ref.~\cite{Cheng:2011ic} 
is most likely the $A$-phase discussed in Ref.~\cite{Aubin:2004dm}, and suggest other ways
to check if indeed this is the case.

We organize this paper as follows. In Sec.~\ref{sec:stag_lag} we define the staggered chiral Lagrangian for an arbitrary number of flavors, $n_f$, and summarize the results of previous work with the notation we will use in this paper. Then in Sec.~\ref{sec:mass_deg}, we find all of the minima of the potential in the revelant region [see \eq{condition}]. We focus on the $A$-phase as that has the features seen in one of the broken phases in Ref.~\cite{Cheng:2011ic}. Finally we conclude in  Sec.~\ref{sec:conc}. We include two appendices where we list the masses in the $A$-phase and the $T$-phase in Appendix~\ref{app:A} and Appendix~\ref{app:T}, respectively. 

\section{The Staggered Chiral Lagrangian}\label{sec:stag_lag}

The starting point of our analysis is the \schpt\
Lagrangian for $n_f$ flavors of quarks \cite{Aubin:2003mg}.
The Lagrangian is written in terms of the field 
$\Sigma=\exp(i\Phi / f)$, a $4n_f \times 4n_f$
matrix, with
\begin{eqnarray}\label{eq:Phi}
  \Phi = \left( \begin{array}{cccc}
    U  & \pi^+ & K^+ & \cdots  \\*
    \pi^- & D & K^0  & \cdots \\*
    K^-  & \bar{K^0}  & S & \cdots\\
    \vdots & \vdots & \vdots & \ddots
  \end{array} \right)\ .
\end{eqnarray}
The elements shown are each $4\times 4$ matrices that are
linear combinations of the hermitian generators,
\begin{equation}\label{eq:T_a}
  T_a = \{ \xi_5, i\xi_{\mu5}, i\xi_{\mu\nu}, \xi_{\mu}, \xi_I\}.
\end{equation}
In euclidean space, the gamma matrices $\xi_{\mu}$ are hermitian, 
and we use the notations
$\xi_{\mu\nu}\equiv \xi_{\mu}\xi_{\nu}$ [$\mu <\nu$ in
\eq{T_a}], $\xi_{\mu5}\equiv \xi_{\mu}\xi_5$ and $\xi_I \equiv
I$ is the $4\times 4$ identity matrix. Under the chiral 
$SU(4n_f)_L\times SU(4n_f)_R$ symmetry, $\Sigma \to L\Sigma
R^{\dagger}$. The components of the diagonal (flavor-neutral)
elements ($U_a$, $D_a$, $S_a$, \etc) are real, while the off-diagonal (flavor-charged)
fields are complex ($\pi^+_a$, $K^0_a$, \etc), such that $\Phi$ is
hermitian. 

From Ref.~\cite{Aubin:2003mg}, the Lagrangian is given by
\begin{eqnarray}\label{eq:final_L}
  \cL & = & \frac{f^2}{8} \Tr(\partial_{\mu}\Sigma 
  \partial_{\mu}\Sigma^{\dagger}) - 
  \frac{1}{4}\mu f^2 \Tr(\cM\Sigma+\cM\Sigma^{\dagger})
	\nonumber\\&&{}
  + \frac{2m_0^2}{3}(U_I + D_I + S_I + \cdots )^2 + a^2 \cV,
\end{eqnarray}
where $\mu$ is a constant with dimensions 
of mass, $f$ is the tree-level pion
decay constant (normalized here so that $f_\pi\approx 131$~MeV), and the $m_0^2$ term includes the $n_f$ flavor-neutral taste-singlet
fields. Normally, in physical calculations, we would take $m_0\to\infty$ 
at the end to decouple the taste-singlet
$\eta'_I$, however in a broken phase there is 
no physical reason to assume a large value for $m_0$, so we will
retain that parameter in our calculations.
Finally, 
$\cV=\cU+\cU\,'$ is the taste-symmetry breaking potential given by
\begin{eqnarray}
  \label{eq:U}
  -\cU  & = & C_1
  \Tr(\xi^{(n_f)}_5\Sigma\xi^{(n_f)}_5\Sigma^{\dagger}) \nonumber \\*
  & & +C_3\frac{1}{2} \sum_{\nu}[ \Tr(\xi^{(n_f)}_{\nu}\Sigma
    \xi^{(n_f)}_{\nu}\Sigma) + h.c.] \nonumber \\*
  & & +C_4\frac{1}{2} \sum_{\nu}[ \Tr(\xi^{(n_f)}_{\nu 5}\Sigma
    \xi^{(n_f)}_{5\nu}\Sigma) + h.c.] \nonumber \\*
  & & +C_6\ \sum_{\mu<\nu} \Tr(\xi^{(n_f)}_{\mu\nu}\Sigma
  \xi^{(n_f)}_{\nu\mu}\Sigma^{\dagger}) \, ,\\
  \label{eq:U_prime}
  -\cU\,'  & = & C_{2V}\frac{1}{4} 
  \sum_{\nu}[ \Tr(\xi^{(n_f)}_{\nu}\Sigma)
    \Tr(\xi^{(n_f)}_{\nu}\Sigma)  + h.c.] \nonumber \\*
  &&+C_{2A}\frac{1}{4} \sum_{\nu}[ \Tr(\xi^{(n_f)}_{\nu
      5}\Sigma)\Tr(\xi^{(n_f)}_{5\nu}\Sigma)  + h.c.] \nonumber \\*
  & & +C_{5V}\frac{1}{2} \sum_{\nu}[ \Tr(\xi^{(n_f)}_{\nu}\Sigma)
    \Tr(\xi^{(n_f)}_{\nu}\Sigma^{\dagger})]\nonumber \\*
  & & +C_{5A}\frac{1}{2} \sum_{\nu}[ \Tr(\xi^{(n_f)}_{\nu5}\Sigma)
    \Tr(\xi^{(n_f)}_{5\nu}\Sigma^{\dagger}) ]\ .
\end{eqnarray}
The $\xi^{(n_f)}_B$ in
$\cV$ are the block-diagonal $4n_f\times 4n_f$ matrices
\begin{equation}\label{eq:xi_B}
  \xi_B^{(n_f)} = 
  \left( \begin{array}{cccc}
    \xi_B & 0 & 0 & \cdots\\*
    0 & \xi_B & 0 & \cdots\\*
    0 & 0 & \xi_B & \cdots\\*
    \vdots & \vdots & \vdots & \ddots
  \end{array} \right),
\end{equation}
with $B \in \{5,\mu5,\mu\nu(\mu <\nu),\mu,I \}$. The mass matrix, $\cM$, is the $4n_f\times 4n_f$ diagonal matrix $\cM = m I_{4n_f\times 4n_f}$, as we are only interested in the degenerate case that is relevant for these BSM studies.

As is well known \cite{Lee:1999zx,Aubin:2003mg}, while
this potential breaks 
the taste symmetry
at $\cO(a^2)$, an accidental $SO(4)$ symmetry remains.
This implies 
a degeneracy in the masses among different tastes of a given flavor
meson, which is seen in the
tree-level masses
of the pseudoscalar mesons. We can classify these mesons into irreducible
representations of $SO(4)$.
The mass for the meson $M$ (composed of quarks $a$
and $b$) with taste $B$, is given at tree-level by\footnote{Note we do not include the
$m_0$ term here for simplicity.}
\begin{equation}\label{eq:tree_lev_mass}
  m^2_{M_B} = 
  \mu \left(m_a + m_b \right) + 
  a^2\Delta_B \ ,
\end{equation}
for mesons composed of different quarks, and 
\begin{equation}\label{eq:tree_lev_mass2}
  m^2_{M_B} = 
  2\mu m_a + 
  a^2\Delta_B + \frac{n_{f} n_t}{4}a^2\delta'_{B-}  \ , \quad  B=V,A\ ,
\end{equation}
for the flavor-neutral mesons. The $\Delta_B$'s are given in Ref.~\cite{Aubin:2003mg} and are linear combinations of the coefficients in the potential $\cU$ and we have the hairpin 
terms,
\begin{equation}\label{eq:mix_vertex_V}
  	\delta'_{V(A)\pm} \equiv \frac{16}{f^2} \left[C_{2V(A)} \pm
    C_{5V(A)}\right] \ .
\end{equation}
The difference in Eq.~(\ref{eq:tree_lev_mass2}) from previous works is that we have the factor $n_fn_t/4$ in front of the hairpin parameter. This arises using the replica method \cite{Bernard:2006zw} to write our expressions for general numbers of flavors and tastes. Of course, $n_f$ is the number of (degenerate) staggered flavors in our calculation, while $n_t$ is the number of tastes per flavor we wish to keep (hence the factor of 1/4). The factor $n_fn_t\equiv n_q$ will be the number of degenerate fermions we have in our theory.\footnote{We note that in our calculations, if $n_t = 4$, then we are not rooting the underlying theory, and as such the theory does not correspond to ``rooted''  staggered quarks.}

Given that empirically the $\Delta_B$'s are all positive in simulations, $\delta'_{V-}$ is consistent with zero, and $\delta'_{A-}<0$ \cite{Aubin:2004fs,Bazavov:2009bb,MILCparams}, focus has been on the possibility of a negative mass-squared arising with the $\eta'_A$ meson. It was shown in Ref.~\cite{Aubin:2004dm} that in current $2+1$-flavor simulations, 
it is very unlikely the simulation will be performed in this phase. Instead, there has been evidence of this phase appearing in BSM simulations \cite{Cheng:2011ic}, and this can easily be understood from Eq.~(\ref{eq:tree_lev_mass}). As discussed in the Introduction, and shown in Figs.~\ref{fig:nqvsmpia} and \ref{fig:nqvsmpib}, assuming that the actual value of $\delta'_{A-}$ is, to a first approximation,
dependent only upon the specific fermion formulation and not the number of flavors (or tastes), then as $n_q$ increases, the simulations are more likely to be performed in the phase described in Ref.~\cite{Aubin:2004dm}.\footnote{As these parameters are non-perturbative low-energy constants, they would have a dependence upon the number of quarks in the simulation, but without knowing that dependence \emph{a priori}, we take them to be independent of $n_q$ as an initial approximation.}

As discussed in Ref.~\cite{Aubin:2004dm}, in the $A$-phase, all of the squared meson masses will 
be positive given the relationships between the different parameters, except possibly for the
tensor taste flavor singlet,  
$\eta'_{ij}$. Specifically, we have (rewriting this expression with our notation),
\begin{eqnarray}\label{eq:tensormass}
  	m^2_{\eta'_{ij}} &=&  
  	-\frac{n_fn_t}{4}a^2\delta_{A-}' - \frac{n_fn_t}{4}a^2\delta_{V+}'
	\nonumber\\&&{}+
  	\frac{16 m^2 \mu^2\left(a^2\Delta_T - a^2\Delta_A + 
    	\frac{n_fn_t}{4} a^2\delta_{V+}'\right )}
       	{\left(a^2\Delta_A + \frac{n_fn_t}{4}a^2\delta_{A-}' \right)^2}.
\end{eqnarray}
The parameter $\delta_{V+}'$ in this expression has not yet been measured, and as such, $m^2_{\eta'_{ij}}$ has the possibility of going negative. This new phase, which we denote the $A'$-phase, could in principle arise in the staggered phase diagram. In the next section we study the phase diagram in general and find that this is not the case, while additional phases other than the $A'$-phase do exist.

\section{General phase diagram}\label{sec:mass_deg}

To find the vacuum state of the theory, 
we must minimize the potential,
\begin{equation}\label{eq:potential}
 \cW =  - \frac{1}{4}\mu m f^2 \Tr(\Sigma+\Sigma^{\dagger})
   + a^2 \cU + a^2 \cU\,' \ ,
\end{equation}
where we have already substituted $\cM = m I_{4n_f\times 4n_f}$.
This calculation is most simply done in the physical basis, where
everything is written in terms of (for three flavors) $\pi^0$, $\eta$, and $\eta'$ instead
of the flavor-basis mesons $U$, $D$, and
$S$. For degenerate quarks, we define the singlet as
\begin{equation}
  \eta'_B  =  \frac{1}{\sqrt{n_f}} \left( U_B + D_B + S_B + \cdots \right)\, ,
\end{equation}
for any number of flavors/tastes. As these are the mesons most likely to acquire a negative mass-squared, we focus solely on these. From here on we remark that in the degenerate quark mass limit, the octet meson masses of a given taste have equal masses which we denote with $m_\pi$,
while the $\eta'$ masses are distinct from these. We note that the number of flavors $n_f$ (so long as it's greater than 1) will not affect our results; $n_f$ will only indicate a greater likelihood of being in the broken phase at this point. 

Generally, $\delta'_{A-}$ and $\delta'_{V-}$ are the parameters likely to be 
negative, and they only arise in the $\eta'$ masses. We infer the 
symmetry breaking to only occur in the $\eta'$ direction in flavor space.  
Therefore, we are going to keep only this meson in our 
expression for $\Phi$ when 
looking for the minima of the potential in \eq{potential}. 
This will be valid right near the critical point, and 
since we are looking at this perturbatively, 
we are looking only at small fluctuations about the minimum.
So long as no other squared mass goes negative in the phase,
our results should give us the correct mass spectrum for the 
broken phase. If a squared mass does go negative, 
as in \eq{tensormass} for certain
values of $\delta'_{V+}$, we are not near a 
minimum of the potential, and thus such additional phases are not stable. 

Keeping only the $\eta'_B$, $\Phi$ and $\Sigma$ are
block-diagonal in flavor space. We can
write the condensate $\langle\Sigma\rangle$ 
in terms of the 16 real numbers $\sigma_I, \sigma_{\mu5}, \sigma_{\mu\nu} (\mu<\nu), \sigma_\mu,$ and $\sigma_5$, 
\begin{equation}
  \langle\Sigma\rangle = \sigma_I \left(I_{4n_f\times 4n_f}\right) 
  + 	i \sigma_{\mu5} \left(i \xi^{(n_f)}_{\mu 5}\right)
  + 	i \sigma_{\mu\nu} \left(i \xi^{(n_f)}_{\mu \nu}\right)
  + 	i \sigma_\mu  \xi^{(n_f)}_{\mu}
  + 	i \sigma_5 \left(i \xi^{(n_f)}_{ 5}\right)\ ,
\end{equation}
with the condition that $\sum_B \sigma_B^2  = 1$. Upon substituting this into the potential, we find the potential (not surprisingly) is only dependent upon the \emph{magnitudes} of these sets of coefficients, given by
\begin{eqnarray}
	\sigma_A & = & \left(\sum_{\mu} \sigma_{\mu5}^2\right)^{1/2} \ ,\\
	\sigma_T & = & \left(\sum_{\mu<\nu} \sigma_{\mu\nu}^2\right)^{1/2} \ ,\\
	\sigma_V & = & \left(\sum_{\mu} \sigma_{\mu}^2\right)^{1/2} \ ,
\end{eqnarray}
so that we have, up to an unimportant constant and for arbitrary numbers of flavors/tastes,
\bea
	\cW
	&=&
	-\frac{3f^2}{2}\biggl[
	4 \mu  m \sigma_I
	-\sigma_A^2 \left(a^2\Delta_A + \frac{n_fn_t}{4}  a^2\delta '_{A-}\right)
	\nonumber
	\\&&{}
	-\sigma_V^2 \left(a^2\Delta_V + \frac{n_fn_t}{4}a^2\delta'_{V-}\right)
	-\sigma_T^2 a^2\Delta_T
	\biggr]\ .
\eea
When minimizing this potential we find three distinct non-trivial phases:
\bea
	A\textrm{-phase} &:& a^2\Delta_A+ \frac{n_f n_t}{4}a^2\delta'_{A-} < -2\mu m \label{eq:conditions1}\ ,\\
	V\textrm{-phase} &:&  a^2\Delta_V + \frac{n_fn_t}{4}a^2\delta'_{V-} < -2\mu m\label{eq:conditions2} \ , \\
	T\textrm{-phase} &:&   a^2\Delta_T < -2\mu m \label{eq:conditions3}\ .
\eea
The $A$-phase was discussed in detail in Ref.~\cite{Aubin:2004dm}, and the results for the $V$-phase are identical to those for the $A$-phase with the replacement $A\leftrightarrow V$ in all of the relevant equations. The $T$-phase is distinct here, and it is unlikely that a simulation will be performed in this phase. This is because all of the parameters $\Delta_B$ are positive in simulations, so these conditions are only likely to hold for the $A$ and $V$ phases because the parameters
$\delta'_{A-}$ and $\delta'_{V-}$ tend to be negative. Nevertheless, we discuss this phase briefly in the Appendix for completeness. We note that in principle, more than one of the conditions in \eqsthru{conditions1}{conditions3} may hold simultaneously, but in fact we only see these three phases. This implies that only one condition will point to the true minimum about which to expand. In the unlikely case that two of the left-hand sides are equal, for example, 
\be
	a^2\Delta_T = a^2\Delta_A+ \frac{n_f n_t}{4}a^2\delta'_{A-} \ ,
\ee
this would introduce a symmetry between (in this case) the axial- and tensor-tastes, but this does not introduce a distinct phase. 

We note that none of these three broken phases correspond to the $A'$-phase discussed in the previous section. Approaching this phase from $A$-phase, we find a saddle point in the potential, and as such this is an unstable equilibrium point. Thus, we will not explore that case further.

We have for the $A$-phase,
\bea
	\sigma_T  = \sigma_V = 0 ,\  \sigma_I\equiv \cos\theta_A = \frac{-2\mu m }{a^2\Delta_A + \frac{n_fn_t}{4} a^2 \delta'_{A-} },\ \sigma_A = \sqrt{1-\sigma_I^2}\ ,
\eea
the $V$-phase,
\bea	
	 \sigma_T  = \sigma_A = 0 ,\   \sigma_I \equiv \cos\theta_V= \frac{-2\mu m }{a^2\Delta_V + \frac{n_fn_t}{4} a^2 \delta'_{V-} },\ \sigma_V = \sqrt{1-\sigma_I^2}\ ,
\eea
and for the $T$-phase,
\bea
	\sigma_A = \sigma_V = 0 ,\  \sigma_I \equiv \cos\theta_T= \frac{-2\mu m }{a^2\Delta_T },\ \sigma_T = \sqrt{1-\sigma_I^2}\ .
\eea
These break the remnant $SO(4)$ symmetry [to $SO(3)$ for the $A$ and $V$ phases, and to $SO(2)\times SO(2)$ for the $T$ phase]. The direction of each of the vectors $a_\mu,$ $a_{\mu5}$, and $a_{\mu\nu}$ is arbitrary, and we will 
choose a particular direction in \eqs{condensate}{TBdirections}.

In each of these cases, we have the condensate of the form
\begin{eqnarray}\label{eq:condensate}
  \left \langle \Sigma   \right \rangle & =&
  \left( \begin{matrix}
     \exp \left[ i \theta_B T_B \right]& 0 &0  & \cdots\\*
    0 & \exp \left[ i \theta_B T_B \right] &  0 & \cdots \\*
    0 &0  & \exp \left[ i \theta_B T_B \right]& \cdots \\*
    \vdots & \vdots & \vdots & \ddots
 \end{matrix}\right) \, ,
\end{eqnarray}
where $B=A,V,T$, and 
\be\label{eq:TBdirections}
	T_B = 
	\begin{cases}
	i\xi_{45} & A-{\rm phase}\\
	\xi_{4} & V-{\rm phase}\\
	i\xi_{12} & T-{\rm phase}
	\end{cases}\ .
\ee

In each of these cases, the shift-symmetry \cite{Lee:1999zx,Aubin:2003mg,Aubin:2004dm} that exists which has the form in the chiral theory,
\be\label{eq:shiftsym}
	\Sigma \to \xi^{(n_f)}_\mu\Sigma  \xi^{(n_f)}_\mu\ ,
\ee
is broken. Thus, as was seen in Ref.~\cite{Cheng:2011ic}, one can use the difference between neighboring plaquettes and the difference between neighboring links to determine if we are in a broken phase.
However, those parameters are sensitive \emph{only} to the breaking of the single-site shift symmetry, so they
cannot distinguish between the $A$, $V$, or $T$-phase. 
Thus, for a complete understanding of the
phase seen in the simulation, the mass spectrum should also be studied.

The key difference between the unbroken phase and the various broken phases is that the squared-meson masses in the broken phases have the generic form
\begin{equation}\label{eq:massesAphase}
	m^2_M = \cA + \cB m^2\ .
\end{equation}
Here $\cA$ and $\cB$ are independent of the quark mass but are dependent upon $a$. Unlike the unbroken phase, the squared meson masses are linear
 in $m^2$ as opposed to $m$, and for some mesons 
 $\cB=0$ so that the mesons have a mass independent of the quark mass. 

\begin{figure}[tbp]
\begin{center}
\includegraphics[width=3in]{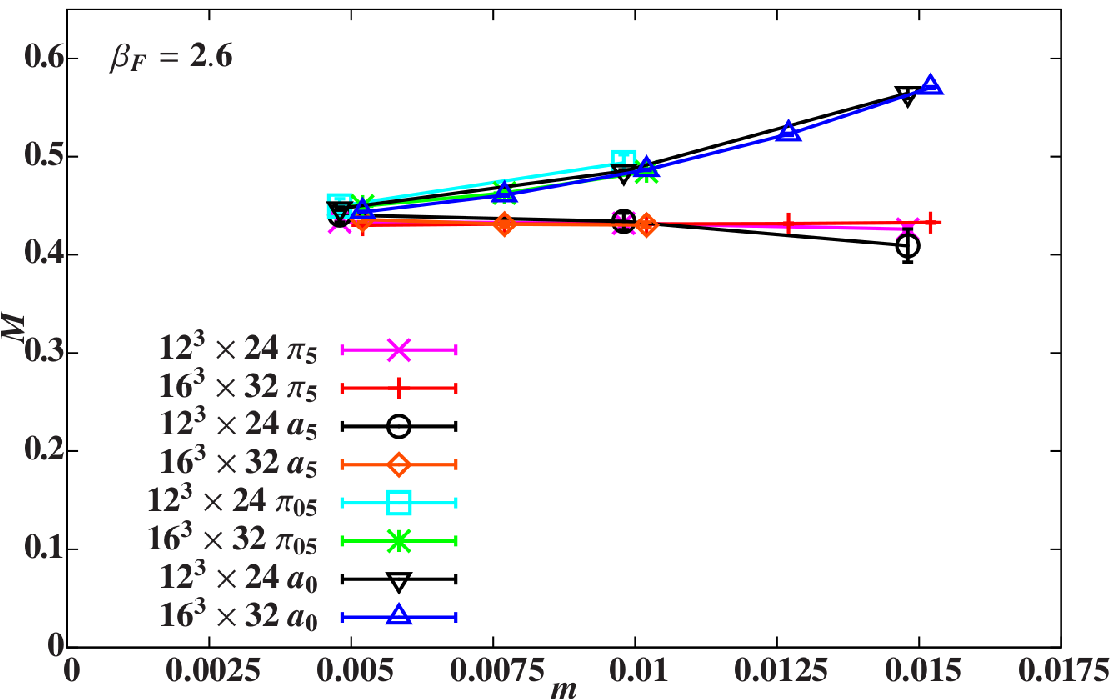}
\includegraphics[width=3in]{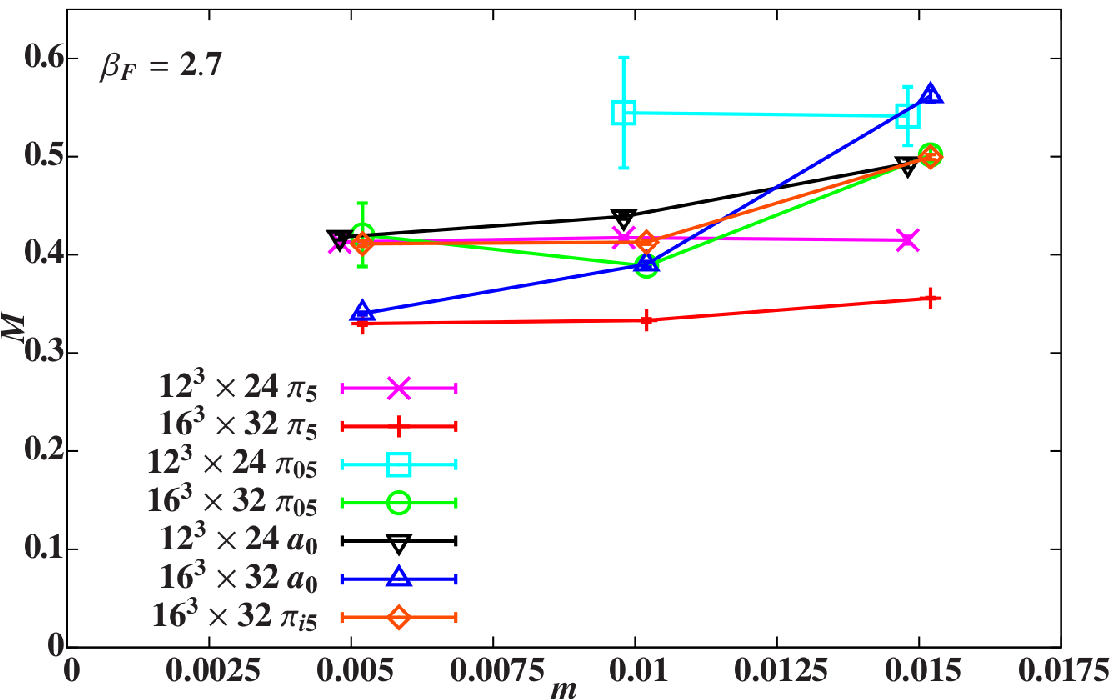}

(a) \hspace{3in} (b)

\caption{Figure 7 from Ref.~\cite{Cheng:2011ic}, showing the masses of the $\pi_5$ and $\pi_{05}$ ($\pi_{45}$ in our notation) as a function of the quark mass. (a) shows the masses in the broken phase discussed here and (b0 shows the masses in the unbroken phase.}
\label{fig:Fig7}
\end{center}
\end{figure}

Figure \ref{fig:Fig7} is a reprint of Fig.~7 from Ref.~\cite{Cheng:2011ic}, which  shows the masses for the pseudoscalar taste pion as well as the taste-45 pion for two lattice spacings.  
Fig.~\ref{fig:Fig7}(a) shows one of the two broken phases seen 
as a function of the input quark mass for 12 quarks (in our notation we would set $n_f=3,n_t=4$). Of the two phase transitions discussed in that paper, the chiral effective theory seems useful for understanding the second here (that appears at smaller lattice spacing around $\beta\approx 2.7$ in Ref.~\cite{Cheng:2011ic}). The authors of Ref.~\cite{Cheng:2011ic} show that the single-site shift symmetry is broken in this phase, and 
we now argue that this is likely the $A$-phase.

The $\pi_5$ in Fig.~\ref{fig:Fig7}(a) has an approximately constant mass as a function of the quark mass, which would be consistent with the calculation of Ref.~\cite{Aubin:2004dm}:
\begin{equation}\label{eq:pi5mass}
	m^2_{\pi_5} =  -\frac{n_f n_t}{4} a^2\delta '_{A-}\ ,
\end{equation}
(recall $\delta '_{A-}<0$ in this phase), and the taste-45 mass has the form,
\begin{equation}\label{eq:pi45mass}
	m^2_{\pi_{45}} =
	m^2_{\pi_{5}}  
	+ a^2\Delta_A \bar{\mu}^2 m^2\ ,
\end{equation}
where $\bar\mu$ is defined below in \eq{mubar}. Were this the 
$V$-phase, the $\pi_5$ would still have a constant mass, but the $\pi_{4}$ would have the behavior of \eq{pi45mass} (with $\Delta_A\to\Delta_V$). Similarly, were this the $T$-phase, as can be seen in Appendix~\ref{app:T}, the $\pi_5$ mass is dependent upon the quark mass while the $\pi_{45}$ mass is constant.

We can see that as $m\to0$, $m_{\pi_{45}}\to m_{\pi_5}$ as r\schpt\ predicts. This gives credence to the fact that the intermediate phase seen in Ref.~\cite{Cheng:2011ic} is in fact this $A$-phase, but a more detailed analysis would require several things. First, one should perform a fit to the forms above for the taste-45 and taste-5 pions. More importantly, one should measure of all of the different taste meson masses to see the pattern as predicted in Ref.~\cite{Aubin:2004dm} (and shown in Appendix~\ref{app:A}). Figure~\ref{fig:Fig7}(b) shows the other side of the transition (larger $\beta$, and thus a smaller lattice spacing), and immediately shows a different pattern: The four axial-taste pions are nearly degenerate and the difference $m_{\pi_{\mu 5}}^2 - m_{\pi_5}^2\approx$ constant as a function of $m$. This is (roughly) the pattern seen in the physical regime of r\schpt~\cite{Lee:1999zx,Aubin:2003mg}.

\begin{figure}[tb]
\begin{center}
\includegraphics[width=4in]{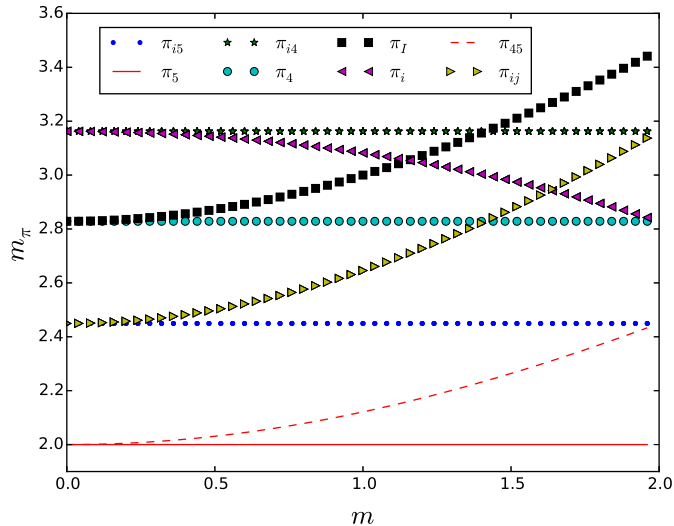}

\caption{$m_\pi$ vs.\ the quark mass in the $A$-phase in arbitrary units. These units are such that when $m\ge2$ the system is in the unbroken phase again. The solid line corresponds to $m_{\pi_5}$ and the dashed line to $m_{\pi_{45}}$. }
\label{fig:mpivsm}
\end{center}
\end{figure}

\begin{figure}[tb]
\begin{center}
\includegraphics[width=4in]{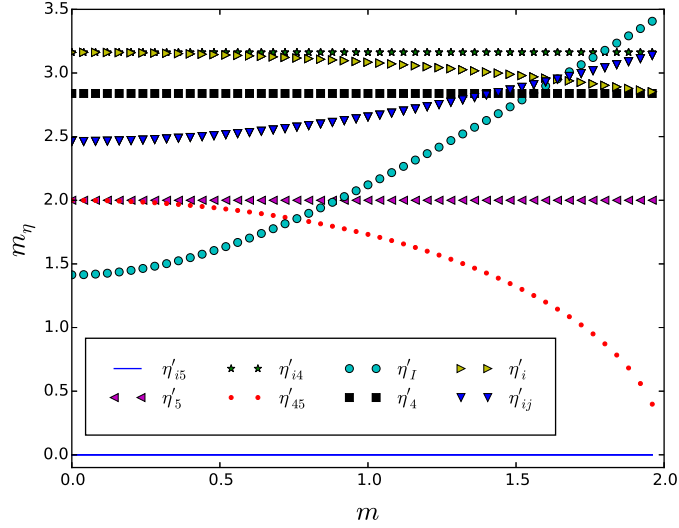}
\caption{$m_\eta$ vs.\ the quark mass in the $A$-phase in arbitrary units. These units are such that when $m\ge2$ the system is in the unbroken phase again. The solid line at $m_\eta =0$ corresponds to the three Goldstone bosons in this phase.}
\label{fig:metavsm}
\end{center}
\end{figure}

We show in Figs.~\ref{fig:mpivsm} and \ref{fig:metavsm} plots of $m_{\pi}$ vs.\ $m$ and $m_{\eta}$ vs.\ $m$ for the different tastes in the $A$-phase. The units in these plots are arbitrary, chosen so that the values $m>2$ correspond to the unbroken phase. The solid red lines in Fig.~\ref{fig:mpivsm} correspond to the $\pi_5$ and $\pi_{45}$ masses, which are to be compared with the masses shown in the left-hand plot of Fig.~7 in Ref.~\cite{Cheng:2011ic}. 

As for the phase at stronger coupling, it is unlikely that r\schpt\ could explain this region. As we have seen 
in this work, r\schpt\ shows that there should be at most four phases: the unbroken phase as well as the $A$, $T$, and $V$ phases. However,
these all are within the regime governed by the constraint in \eq{condition}, most importantly that $a^2 \Lambda^2_{\rm QCD} \ll 1$. The stronger coupling phase likely violates this
constraint, and thus the chiral theory is not valid in this regime. Nevertheless, it would be instructive to understand more about the intermediate phase to be sure that r\schpt\ is in fact 
describing the region as we expect it is.

\section{Conclusion}\label{sec:conc}

We have studied the complete phase diagram for staggered quarks with an arbitrary number of degenerate flavors, at least within the window given in \eq{condition}. In this regime, there are only four phases: one unbroken (physical) phase, as well as three phases where the (approximate) $\cO(a^2)$ accidental $SO(4)$ symmetry is broken. Of these three phases, as seen previously \cite{Aubin:2004dm}, only one phase (the $A$-phase) is likely to be seen in simulations, but only if one looks at theories with 8 or 12 flavors of quarks \cite{Cheng:2011ic}. The additional possible phase that was suggested to exist in Ref.~\cite{Aubin:2004dm} [when the squared mass of the $\eta'_{ij}$, \eq{tensormass}, goes negative] does not correspond to a stable region. 

While r\schpt\ cannot fully explain both broken phases seen in Ref.~\cite{Cheng:2011ic}, it can give a picture of broken phases that are
located close to the unbroken phase (as a function of the lattice spacing). Studying the 
plaquette \& link differences as well as as the staggered meson mass spectrum would allow
one to determine specifically which region of the phase diagram the simulation is in. For BSM studies this is essential as it is more likely to enter these unphysical regimes for additional quark flavors.

\appendix

\section{$A$-Phase}\label{app:A}

In this appendix we list the meson masses that appear in the $A$-phase \cite{Aubin:2004dm}. Here we put them in terms of $n_f n_t$ as above, and we define
\be\label{eq:mubar}
	\overline{\mu} \equiv \frac{-2\mu}{a^2\Delta_A + \frac{n_f n_t}{4} a^2\delta'_{A-}}\ .
\ee
The first masses we list are constant in the quark mass:
\bea
	m^2_{\eta'_{i5}} & = & 0\ ,
	\\
	m^2_{\eta'_{5}} =  m^2_{\pi_{5}} & = & -a^2\Delta_A - \frac{n_f n_t}{4}a^2\delta'_{A-}\ ,
	\\
	m^2_{\pi_{i5}} & = & -\frac{n_f n_t}{4} a^2\delta '_{A-}\ ,
	\\
	m^2_{\pi_{i4}} = m^2_{\eta'_{i4}}   & = &  m^2_{\pi_{5}} + a^2\Delta_T\ ,
	\\
	m^2_{\eta'_{4}} & = &m^2_{\pi_{5}} +a^2\Delta_V 
	+ \frac{n_f n_t}{4}a^2\delta'_{V-}\ ,
	\label{eq:eq26}
	\\
	m^2_{\pi_{4}} & = & m^2_{\pi_{5}} +a^2 \Delta_V\ .
\eea
We note that Eq.~(26) in Ref.~\cite{Aubin:2004dm} [corresponding to our \eq{eq26}] has a typo, as the final term in that expression should be $+\frac34 a^2\delta'_V$ in that paper's notation, not $-\frac34 a^2\delta'_V$.
With the above, we can determine the constants,
\[
	\Delta_A, 
	\quad
	\delta'_{A-},
	\quad
	 \Delta_T, 
	\quad
 	\Delta_V ,
	\quad
	\delta'_{V-}  \ .
\]
Then
\bea
	m^2_{\pi_{45}} & = & 
	m^2_{\pi_{5}}  
	+ a^2\Delta_A \bar{\mu}^2 m^2
\eea
allows us to determine $\bar\mu$.
With 
\bea
	m^2_{\pi_{I}} & = & m^2_{\pi_{5}} 
	+ \bar{\mu}^2 m^2 \left(a^2\Delta_I - a^2\Delta_V\right) + a^2\Delta_V
	\\
	m^2_{\eta'_{I}} & = & 
	n_t m_0^2 + 
	m^2_{\pi_{5}} -\frac{n_f n_t}{4} a^2\delta '_{A+}+ a^2\Delta_V
	\nonumber
	\\&&{}
	+\bar{\mu}^2 m^2 \biggl(\frac{n_f n_t}{4} a^2\delta '_{A+} + a^2\Delta_I 
	- a^2\Delta_V\biggr) 
\eea
allows us to determine $\Delta_I$ and $\delta'_{A+}$ respectively,\footnote{Note that if we take the $m_0\to\infty$ limit seriously we would not examine the $\eta'_I$ mass. However, given that we are in an unphysical phase, there is no reason to assume that this is the case, so we keep this mass in our theory. This expression would allow us to obtain $\delta'_{A+}$ along with $m_0$ at the same time as they have different dependencies on the quark 
mass.} and finally
\bea
	m_{\eta'_{ij}}^2 
	&=&
	 \bar{\mu}^2 m^2 m^2_{\pi_{i4}} 
	-\frac{n_f n_t}{4} \left(1- \bar{\mu}^2 m^2\right) (a^2\delta'_{A-} + a^2\delta'_{V+})
\eea
for $\delta'_{V+}$. 

The following four masses are then determined from those above results, 
\bea	
	m^2_{\pi_{i}} & = & 
	m^2_{\pi_{5}} 
	+(m^2_{\pi_{4}}-m^2_{\pi_{5}}) \bar{\mu}^2 m^2 
	+(m^2_{\eta'_{i4}} - m^2_{\pi_{5}})(1 - \bar{\mu}^2 m^2 )\ ,
\\
	m^2_{\pi_{ij}} & = & 
	m^2_{\pi_{i5}} 
	+
	\bar{\mu}^2 m^2 \left(m^2_{\eta'_{i4}}  - m^2_{\pi_{i5}} \right) \ ,
	\\
	m^2_{\eta'_{45}} & = & m^2_{\pi_{5}}  \left(1 - \bar{\mu}^2 m^2\right)\ ,
	\\
	m^2_{\eta'_{i}} & = & 	m^2_{\eta'_{i4}}
	 -  ( m^2_{\eta'_{i4}}
	-  m^2_{\eta'_{4}}) \bar{\mu}^2 m^2\ .
\eea
This shows that we have non-trivial relationships between the various masses. Additionally, as seen in Figs.~\ref{fig:mpivsm} and \ref{fig:metavsm} we have several crossings of the meson masses for both the $\pi$ and the $\eta$. While those figures are for a specific set of parameters, they are indicative of the qualitative features of the $A$-phase. 
	
\section{$T$-Phase}\label{app:T}

In this appendix we list the masses for the mesons in the $T$-phase, where just as before, the octet masses are equal. In this case we define
\be
	\bar \mu \equiv \frac{ 2\mu}{-a^2\Delta_T}\ .
\ee
Thus we have
\bea
	 m^2_{\pi_5} = m^2_{\eta'_5}&=&-a^2\Delta_T \bar \mu^2 m^2
\\
	 m^2_{\pi_I } & = &
	  \bar \mu^2 m^2 \left(a^2\Delta_I-a^2\Delta _T\right)
	  \\
	 m^2_{\eta'_I}&=&
	n_t m_0^2 + 
	  \bar \mu^2 m^2 \left(a^2\Delta_I-a^2\Delta _T\right)
\\
	m^2_{\pi_{12} } =
	m^2_{\eta'_{12}}&=&-a^2\Delta_T (1-\bar \mu^2 m^2)
\\
	m^2_{\pi_{34} } = 
	m^2_{\eta'_{34}}&=&\left(a^2\Delta_I-a^2\Delta _T\right)(1-\bar \mu^2 m^2)
\\
	m^2_{\eta'_1}=
	m^2_{\eta'_2} & =  & a^2\Delta _V  - a^2 \Delta _T+ \frac{n_fn_t}{4} a^2\delta '_{V-}
\\
	m^2_{\eta'_3}=
	m^2_{\eta'_4}&=&
	a^2\Delta _V-a^2\Delta _T	  -\frac{n_fn_t}{4}(1-\bar\mu^2 m^2) a^2\delta '_{A+}
	\nonumber\\&&{}
	+\frac{n_fn_t }{4}\bar \mu^2 m^2 a^2\delta '_{V-}
\\
	m^2_{\eta'_{15}}=
	m^2_{\eta'_{25}} & = & a^2\Delta _A - a^2\Delta _T + \frac{n_fn_t}{4} a^2\delta '_{A-}
\\
	m^2_{\eta'_{35}}=
	m^2_{\eta'_{45}} & =  & 
	a^2\Delta _A   - a^2\Delta _T
	- \frac{n_fn_t}{4} a^2\delta '_{V+}
	\nonumber\\&&{}
	+ \frac{n_fn_t}{4} \bar\mu ^2 m^2 \left(a^2\delta '_{A-} 
	+ a^2\delta '_{V+}\right)
\\
	 m^2_{\pi_{\mu5} }&= & a^2\Delta _A - a^2\Delta _T \\
	 m^2_{\pi_\mu }&= &a^2 \Delta _V - a^2\Delta _T \\
	m^2_{\eta'_{24}}=
	m^2_{\eta'_{23}}=
	m^2_{\eta'_{14}}=
	m^2_{\eta'_{13}}&=&0\\
 m^2_{\pi_{24} }=
 m^2_{\pi_{23} }= 
 m^2_{\pi_{14} }= 
 m^2_{\pi_{13} }&= & 0 
\eea
With $\Delta_T$ ``large and negative,'' these are all positive or zero with the exception of those with the $\delta'_{A+}$ or $\delta'_{V+}$ terms. As in the $A$ or $V$ phase, if those parameters are such that $m^2_{\eta'_{45}}<0$ (for example), this leads to a phase that would not give rise to a stable ground state. 

The generic dependence $m_\pi^2  = \cA + \cB m^2$ persists in this phase, but for one we see a different pattern than in the $A$ or $V$ phases. Additionally, there are no mixings between different taste mesons. Nevertheless, it is unlikely, given the empirical evidence, that one would be able to run a simulation in this phase, and as such we will not discuss this phase further.

\bigskip
\bigskip
\centerline{\bf ACKNOWLEDGMENTS}
\bigskip

We would like to thank Claude Bernard, Anna Hasenfratz, and Steve Sharpe for useful discussions. Additionally we would like to thank Claude Bernard for helpful comments on the manuscript.

\end{document}